\documentclass[seceq,fleqn]{ptptex}

\usepackage{graphicx}

\title{
Magnetic and axial-vector transitions of the baryon  antidecuplet
}

\author{
Hyun-Chul \textsc{Kim}$^{1,}$\footnote{ e-mail address:
hchkim@pusan.ac.kr},
Ghil-Seok \textsc{Yang}$^{2,}$\footnote{ e-mail address:
yang@tp2.rub.de}
and
Klaus \textsc{Goeke}$^{2,}$\footnote{ e-mail address:
goeke@tp2.rub.de} 
}

\inst{
$^1$ Department of Physics, and Nuclear Physics \&
Radiation Technology Institute (NuRI), Pusan National University, 609-735
Busan, Republic of Korea\\
$^{2}$ Institut f\"ur Theoretische Physik II, Ruhr-Universit\" at Bochum,
D--44780 Bochum, Germany
}

\abst{
We report the recent results of the magnetic transitions and
axial-vector transitions of the baryon antidecuplet within the
framework of the chiral quark-soliton model.  The dynamical model
parameters are fixed by experimental data for the magnetic moments of  
the baryon octet, for the hyperon semileptonic decay constants, and 
for the singlet axial-vector constant.  The transition magnetic
moments $\mu_{\Lambda\Sigma}$ and $\mu_{N\Delta}$ are well reproduced
and other octet-decuplet and octet-antidecuplet transitions are
predicted.  In particular, the present calculation of
$\mu_{\Sigma\Sigma^*}$ is found to be below the upper bound
$0.82\mu_N$ that the SELEX collaboration measured very recently.  The
results explains consistently the recent findings of a new $N^*$
resonance from the GRAAL and Tohoku LNS group.  We also obtain the
transition axial-vector constants for the $\Theta^+\to KN$ from which 
the decay width of the $\Theta^{+}$ pentaquark baryon is determined as
a function of the pion-nucleon sigma term $\Sigma_{\pi N}$.  We   
investigate the dependence of the decay width of the $\Theta^{+}$
on the $g_{A}^{(0)}$, with the $g_{A}^{(0)}$ varied within the range
of the experimental uncertainty.  We show that a small decay width
of the $\Theta^{+}\to KN$, i.e. $\Gamma_{\Theta KN} \leq 1$~MeV, is
compatible with the values of all known semileptonic decays with the
generally accepted value of $g_{A}^{(0)} \approx 0.3$ for the proton.  
}

\begin{document}

\maketitle

\section{Introduction}
We would like to begin the present talk by mentioning about two
Japanese physicists who were Yukawa's students and collaborators in 1940's.  
Shoichi Sakata and Mitsuo Taketani developed the three-stages theory
based on a Marxist dialectical philosophy to explain how science is
advanced.   
\vspace{0.3cm}

{\em ``Suppose a researcher discovers a new, inexplicable
  phenomenon.  First he or she learns the details and tries to discern
  regularities.  Next the scientist comes up with a qualitative model
  to explain the patterns and finally developes a precise mathematical
  theory that subsumes the model.  But another discovery soon forces
  the process to repeat.  As a result, the history of science
  resembles a spiral, going round in circles yet always
  advancing.''}~\cite{BrownNambu} 

\vspace{0.3cm}

Since the first experimental observation of a signal of the 
pentaquark baryon $\Theta^+$ by the LEPS collaboration at
SPring-8~\cite{Nakano:2003qx} which was triggered by the theoretical
predictions from the chiral soliton model
($\chi$SM)~\cite{Diakonov:1997mm}, there has been a great amount of 
experimental and theoretical efforts to understand the nature of this
new resonance $\Theta^+$.  However, the null results of the recent
CLAS experiment have cast doubt on its
existence~\cite{Battaglieri:2005er}.  Meanwhile, the DIANA collaboration
has continued to search for the $\Theta^{+}$ and announced very
recently the formation of a narrow $pK^{0\text{ }}$ peak with mass of
$1537\pm2$ MeV/$c^{2}$ and width of $\Gamma=0.36\pm0.11$ MeV in the
$K^{+}n\rightarrow K^{0}p$ reaction~\cite{Barmin:2006we}.  Moreover, 
several new experiments for the $\Theta^{+}$ are under
way~\cite{Hotta:2005rh,Miwa:2006if,Nakano}.  In the present cloudy 
status for the $\Theta^+$, more efforts are required
for understanding the $\Theta^{+}$ theoretically as well as
experimentally.  No matter whether the $\Theta^+$ exists or not, such
efforts will surely not end up in vain, as Sakata and Taketani
put forward~\cite{BrownNambu}.   

In addition to the $\Theta^+$, the GRAAL
collaboration~\cite{Kuznetsov:2004gy,Kuznetsov:2006kt} and the Tohoku
LNS group~\cite{tohoku} announced the evidence of a new nucleon-like 
resonance with a narrow decay width $\sim10$ MeV and a mass 
$\sim1675$ MeV in the $\eta$-photoproduction from the neutron
target.  This new nucleon-like resonance, $N^{\ast}(1675)$, may be
regarded as a non-strange pentaquark because of its narrow decay width
and dominant excitation on the neutron target which are known to be
characteristic for typical pentaquark baryons~\cite{Polyakov:2003dx}. 
Moreover, several theoretical calculations of the $\gamma
N\rightarrow\eta N$ reaction~\cite{Choi:2005ki,Fix:2007st} support the
identification of the $N^{\ast}(1675)$ as a member of the baryon
antidecuplet, based on the values of the transition magnetic moments
in Refs.~\cite{Kim:2005gz,Azimov:2005jj}.

In the present talk, we would like to report recent investigations
on the magnetic and axial-vector transitions of the baryon
antidecuplet within the framework of the chiral quark-soliton
model~\cite{Kim:2005gz,Yang:2007yj}, emphasizing in particular the
$\Theta^+$ and $N_{\overline{10}}^*$.  We include the effects of
flavor SU(3) symmetry breaking and employ the
\textquotedblleft\emph{model-independent
  approach}\textquotedblright~\cite{Adkins:1984cf} in which all 
dynamical parameters of the model are fixed by existing experimental
data.         

\section{Formalism}
The electromagnetic and axial-vector transition form factors are
defined by the following transition matrix elements of the vector and
axial-vector currents~\cite{Kim:2005gz}: 
\begin{eqnarray}
\langle B|V_{\mu}^{Q}|B\rangle&=&\bar{u}_{B}(p_{2})\left[ 
F_{1}(q^{2})\gamma_{\mu}+\frac{iF_{2}(q^{2})}{M_{1}}\sigma_{\mu\nu}q^{\nu
}+\frac{F_{3}(q^{2})}{M_{1}}q_{\mu}\right]  u_{B}(p_{1}),\cr\langle
B_{2}|A_{\mu}^{X}|B_{1}\rangle&=&\bar{u}_{B_{2}}(p_{2})\left[  g_{1}
(q^{2})\gamma_{\mu}-\frac{ig_{2}(q^{2})}{M_{1}}\sigma_{\mu\nu}q^{\nu}
+\frac{g_{3}(q^{2})}{M_{1}}q_{\mu}\right]  \gamma_{5}u_{B_{1}}(p_{1}),
\end{eqnarray}
where the vector and axial-vector currents are defined as
\begin{equation}
V_{\mu}^{Q}\;=\;\bar{\psi}(x)\gamma_{\mu}\lambda_{Q}\psi(x),\;\;\;
A_{\mu}^{X}\;=\;\bar{\psi}(x)\gamma_{\mu}\gamma_{5}\lambda_{X}\psi(x)
\label{Eq:current} 
\end{equation}
with $Q=\frac{1}{2}(3 + 8/\sqrt{3})$ for the electromagnetic currents
and $X=\frac{1}{2}(1\pm i2)$ for strangeness conserving $\Delta S=0$
currents and $X=\frac{1}{2}(4\pm i5)$ for $|\Delta S|=1$. The
$q^{2}=-Q^{2}$ stands for the square of the momentum transfer
$q=p_{2}-p_{1}$. The form factors $F_{i}$ and $g_{i}$ are real
quantities due to $CP$-invariance, depending only on the square of the
momentum transfer, among which we are mostly interested in $F_2$ and
$g_1$.  Taking into account the $1/N_{c}$ 
rotational and $m_{\mathrm{s} }$ corrections, we can write the
resulting magnetic moments $\mu^{(B)}$ and axial-vector constants 
$g_{1}^{(B_{1}\rightarrow B_{2})}(0)$ as follows: 
\begin{eqnarray}
\mu^{(B)} &=& w_{1}\langle
B|D_{Q3}^{(8)}|B\rangle\;+\;w_{2}d_{pq3}\langle
B|D_{Qp}^{(8)}\,\hat{J}_{q}
|B\rangle\;+\;\frac{a_{3}}{\sqrt{3}}\langle B|D_{Q8}^{(8)}\,\hat 
{J}_{3}|B\rangle\cr
&+& m_{s}\left[  \frac{w_{4}}{\sqrt{3}}d_{pq3}\langle B|D_{Qp}
^{(8)}\,D_{8q}^{(8)}|B\rangle+w_{5}\langle B|\left(  D_{Q3}
^{(8)}\,D_{88}^{(8)}+D_{Q8}^{(8)}\,D_{83}^{(8)}\right)
|B\rangle\right. \nonumber\\
&&  +\left.  w_{6}\langle B|\left(D_{Q3}^{(8)}\,D_{88}^{(8)}-D_{Q8} 
^{(8)}\,D_{83}^{(8)}\right)  |B\rangle\right],\cr
g_{1}^{(B_{1}\rightarrow B_{2})}(0) & = & a_{1}\langle B_{2}|D_{X3}^{(8)}
|B_{1}\rangle\;+\;a_{2}d_{pq3}\langle B_{2}|D_{Xp}^{(8)}\,\hat{J}_{q}
|B_{1}\rangle\;+\;\frac{a_{3}}{\sqrt{3}}\langle B_{2}|D_{X8}^{(8)}\,\hat
{J}_{3}|B_{1}\rangle\cr
&+& m_{s}\left[  \frac{a_{4}}{\sqrt{3}}d_{pq3}\langle B_{2}|D_{Xp}
^{(8)}\,D_{8q}^{(8)}|B_{1}\rangle+a_{5}\langle B_{2}|\left(  D_{X3}
^{(8)}\,D_{88}^{(8)}+D_{X8}^{(8)}\,D_{83}^{(8)}\right)  |B_{1}\rangle\right.
\nonumber\\
&&  +\left.  a_{6}\langle B_{2}|\left(  D_{X3}^{(8)}\,D_{88}^{(8)}-D_{X8}
^{(8)}\,D_{83}^{(8)}\right)  |B_{1}\rangle\right]  ,\label{Eq:g1}
\end{eqnarray}
where $w_i$ and $a_{i}$ denote parameters encoding the specific
dynamics of the chiral quark-soliton model.  $\hat{J}_{q}$
($\hat{J}_{3}$) stand for the $q$-th (third) components of the
collective spin operator of the baryons, respectively.  The
$D_{ab}^{(\mathcal{R})}$ denote the SU(3) Wigner matrices in
representation $\mathcal{R}$.  The dynamical parameters $w_i$ and
$a_i$ are fixed by the experimental data for the baryon magnetic
moments, and hyperon semileptonic decay constants and singlet
axial-vector constant of the proton, respectively.  The magnetic
moments of the baryon decuplet have been obtained in Ref.~\cite{Yang},
where those of the $\Delta$ and $\Omega$ baryons were well reproduced.
Having obtained the numerical results of $w_i$ and $a_i$, we can
immediately obtain the transition magnetic moments and axial-vector
constants from the baryon octet to the antidecuplet.   

\section{Results and discussion}
\begin{figure}[h]
\begin{center}
\includegraphics[scale=0.42]{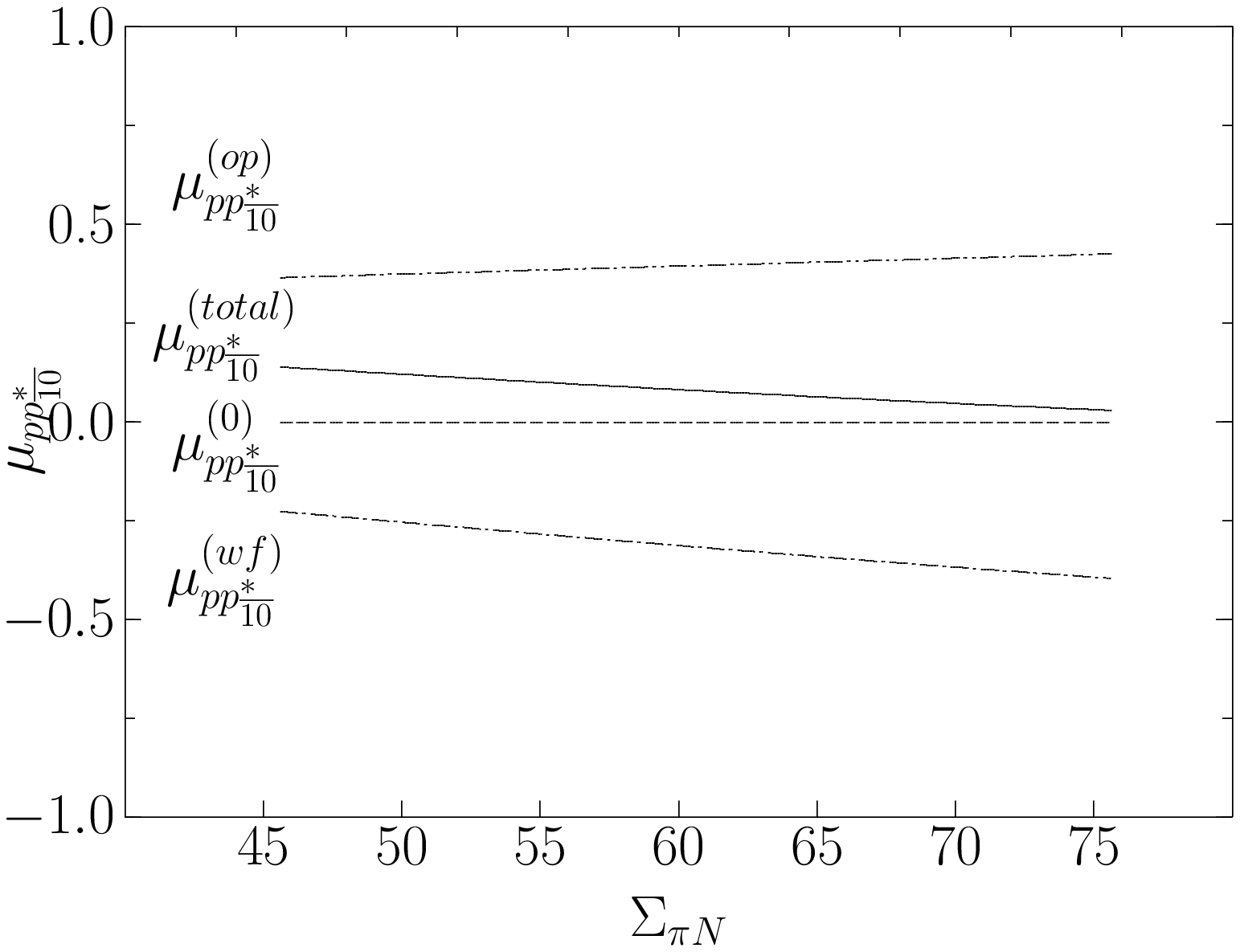} \hspace{0.3cm}
\includegraphics[scale=0.42]{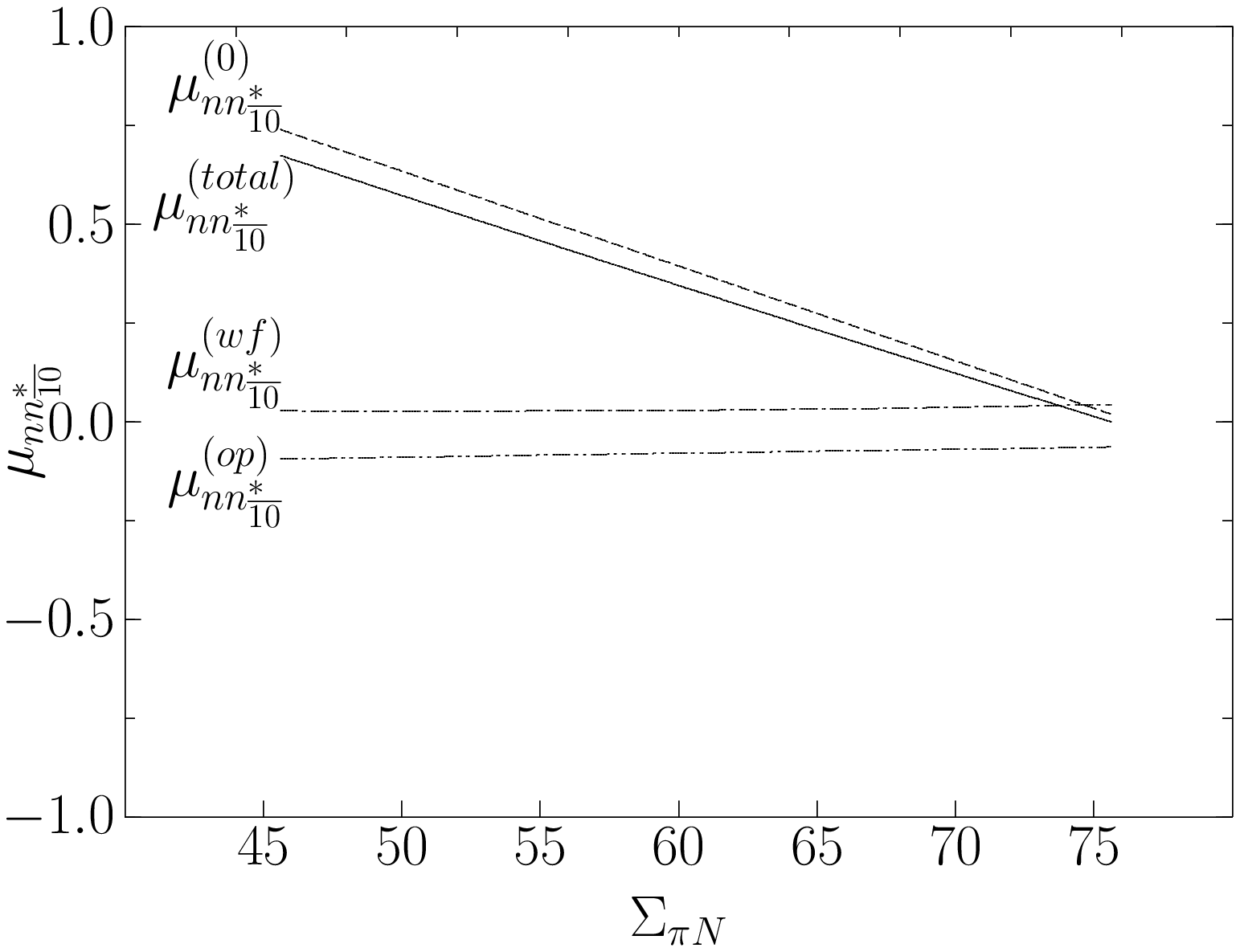}
\end{center}
\caption{$p^{\ast}\rightarrow p$ transition magnetic
moment as functions of $\Sigma_{\pi N}$ in the left panel and
$n^{\ast}\rightarrow n$ one as functions of $\Sigma_{\pi N}$ in the
right panel.}
\label{fig:1}
\end{figure}

In Table~\ref{tab1}, we list the numerical results for the transition 
magnetic moments of the nonexotic and exotic baryons for three
different values of the $\Sigma_{\pi N}$ in units of $\mu_N$. 
\begin{table}[h]
\begin{tabular}{c|cccccccc}
\hline\hline
$\Sigma_{\pi N}$ [MeV] & $\mu_{N\Delta}$ & $\mu_{\Lambda^0\Sigma^0}$ & $
\mu_{\Sigma^{-}\Sigma^{\ast-}}$ & $\mu_{pp_{\overline{10}}^{\ast}}$ &
$\mu_{nn_{\overline{10}}^{\ast}}$ &
$\Gamma_{pp_{\overline{10}}^{\ast}}$ &
$\Gamma_{nn_{\overline{10}}^{\ast}}$ & $\Gamma_{nn_{\overline{10} 
}^{\ast}}/\Gamma_{pp_{\overline{10}}^{\ast}}$ \\ \hline
$50$ & $-3.06$ & $1.54$ & $-0.44$ & $0.12$ & $0.56$ & $11.5$ & $250$ &
$21.67$ \\  
$60$ & $-3.16$ & $1.58$ & $-0.50$ & $0.08$ & $0.33$ & $5.12$ & $87.2$
& $17.02$ \\ 
$70$ & $-3.31$ & $1.64$ & $-0.59$ & $0.04$ & $0.11$ & $1.28$ & $9.69$
& $7.56$ \\ \hline\hline 
\end{tabular}%
\caption{Transition magnetic moments in units of $\protect\mu_N$. The
experimental value for $\protect\mu_{\Lambda^0\Sigma^0}$ is given as
$(1.61\pm0.08)\,\protect\mu_N$. The empirical value for
$|\protect\mu_{N\Delta}|$ is approximately equal to $3.1\,\protect 
\mu_N$.  Partial decay widths $\Gamma_{NN_{\overline{10}}^{\ast}}$
for the radiative decays of exotic and nonexotic baryons in units of
keV. The last column stands for the ratio of the partial decay widths
$n_{\overline{10}}^{\ast}\rightarrow n+\protect \gamma$ and
$p_{\overline{10}}^{\ast}\rightarrow p+\protect\gamma$.} 
\label{tab1}
\end{table}
Those for $\mu_{N\Delta}$ and $\mu_{\Lambda^0\Sigma^0}$ are in a very
good agreement with the experimental and empirical
data~\cite{Kim:2005gz}.  The upper bound for
$|\mu_{\Sigma^{-}\Sigma^{\ast-}}|$ extracted from the upper limit for
the partial decay width of the SELEX experiment is 
around $0.82\,\mu_N$~\cite{Molchanov:2004iq}.  The present prediction 
for $\mu_{\Sigma^{-}\Sigma^{\ast-}}$ lies definitely in the allowed
region for all reasonable values of $\Sigma_{\pi N}$.  The
$N_{\overline{10}}^*\to N$ transition magnetic moments are rather
sensitive to the $\Sigma_{\pi N}$.  It is very similar
to the case of the magnetic moments of the baryon
antidecuplet~\cite{Yang}. The reason is due to the fact that 
$\mu_{NN_{\overline{10}}^{\ast}}$ is proportional to
$w_{1}+w_{2}+\frac{1}{2}w_{3}$, so that the terms with the
$\Sigma_{\pi N}$ in $w_1$ and $w_2$ interfere linearly.  As a
result, $\mu_{nn_{\overline{10}}^{\ast}}$ decreases monotonically as
$\Sigma_{\pi N}$ increases, as shown in Fig.~\ref{fig:1}.  

Since the $p^{\ast}\rightarrow p$
transition one is SU(3) forbidden, it is rather small.  More
interestingly, the neutron transition magnetic moment is about three
to five times larger than the proton one.  The partial width of
radiative decays from the baryon antidecuplet to the octet is
expressed as  
\begin{equation}
\Gamma(B_{\overline{10}}\to B_8\gamma) = 4 \alpha_{\rm EM}
\frac{E_\gamma^3 }{(M_8 + M_{\overline{10}})^2}
\left(\frac{{\mu}_{B_8B_{\overline{10}}}}{\mu_N}\right)^2,  
\end{equation}
where $\alpha_{\rm EM}$ denotes the fine structure constant and
$E_\gamma$ is the energy of the produced photon:
\begin{equation}
E_\gamma = \frac{M_{10 (\overline{10})}^2-M_8^2}{2M_{10
    (\overline{10})}}.
\label{eq:radcay}
\end{equation} 
Since the partial decay width $\Gamma_{nn_{\overline{10}}^{\ast}}$ is
proportional to the transition magnetic moment as shown in
Eq.(\ref{eq:radcay}), it turns out to be about eight to twenty times
larger than $\Gamma_{pp_{\overline{10}}^{\ast}}$.  This result is
consistent with those of the GRAAL and Tohoku
experiments~\cite{Kuznetsov:2006kt,tohoku}.    

\begin{figure}[ht]
\centering
\includegraphics[scale=0.6]{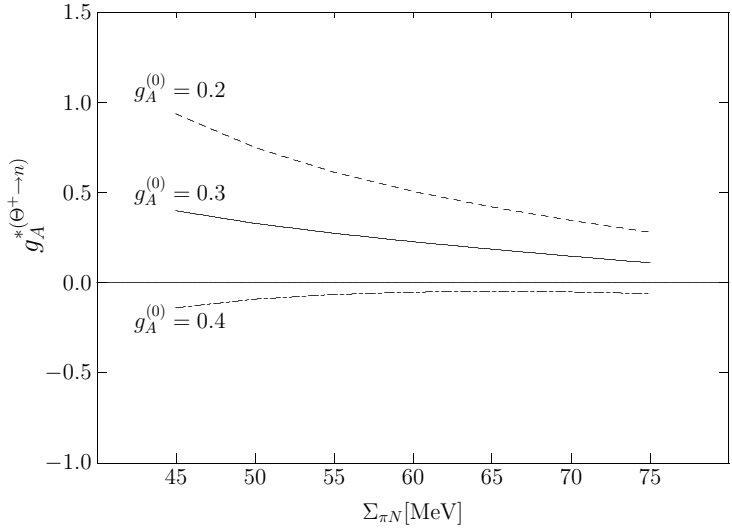}
\caption{The transition axial-vector coupling constant for
  $\Theta^+\to K^{+} n$ as a function of $\Sigma_{\pi N}$.  The solid
  curve denotes that with $g_A^{(0)}=0.3$,
  while the dashed and dot-dashed ones represent that with
  $g_A^{(0)}=0.2,\;0.4$, respectively.}
\label{fig:2}
\end{figure}
In Fig.~\ref{fig:2}, the transition axial-vector coupling constant for 
$\Theta^+\to K^{+} n$ is drawn as a function of $\Sigma_{\pi N}$,
three different values of the singlet axial-vector constant
$g_A^{(0)}$ for the proton being varied from 0.2 to 0.4.  The larger
$g_A^{(0)}$ we use, the smaller $g_A^{*(\Theta\to n)}$ we obtain.
When $g_A^{(0)}$ is larger than $0.37$, the $g_A^{*(\Theta\to n)}$
becomes even negative.  

Since the decay width of the $\Theta^+\to KN$ is proportional to the
square of the transition axial-vector constant as follows:
\begin{equation}
\Gamma_{\Theta KN}  = 2\Gamma_{\Theta K^{+} n} =
 \frac{\left(g_{A}^{\ast(\Theta\to n)}\right)^{2} |\vec{p}|}{16\pi
 f_{K}^{2}M_{\Theta}^{2}}\left[\left(M_{\Theta}-M_{N}
\right)^{2}-m_{K}^{2}\right]\left(M_{\Theta}+M_{N}\right)^{2},   
\end{equation}
it is rather sensitive
to the $g_A^{*(\Theta\to n)}$.  In Table~\ref{tab:results}, we list 
the numerical results of the the decay width of the $\Theta^+\to KN$,
$\Gamma_{\Theta KN}^{\mathrm{total}}$ for four different values of the
$g_A^{(0)}$ as a function of $\Sigma_{\pi N}$.  The $\Gamma_{\Theta
  KN}^{\mathrm{total}}$ turns out to be the smallest with 
$g_A^{(0)}=0.36$ and $\Sigma_{\pi N}$. 
\begin{table}[ht]
\begin{tabular}{c|cccc}
\hline\hline
\multicolumn{1}{c|}{$\Gamma_{\Theta KN}^{({\rm total})}$}&
\multicolumn{4}{c}{Input $g_{A}^{(0)}$}\\
\textbf{\small $\Sigma_{\pi N}[{\rm MeV}]$}& \textbf{\small
~~~~$0.28$~~~~}& \textbf{\small ~~~~$0.32$~~~~}& \textbf{\small
~~~~$0.36$~~~~}& \textbf{\small ~~~~$0.40$~~~~} \\ \hline 
$50$& $22.25$& $7.82$& $0.76$&
$1.10$\\
$60$& $10.45$& $3.82$& $0.46$&
$0.36$\\
$70$& $4.54$& $1.50$& $0.10$&
$0.35$\\
\hline\hline
\end{tabular}
\caption{The decay width of $\Theta^+\to K N$ determined with
$g_{A}^{(0)}$ varied from $0.28$ to $0.40$. The $\Sigma_{\pi N}$
is varied from $45$ to $75$ MeV.} \label{tab:results}
\end{table}
\begin{figure}[ht]
\centering
\includegraphics[scale=0.6]{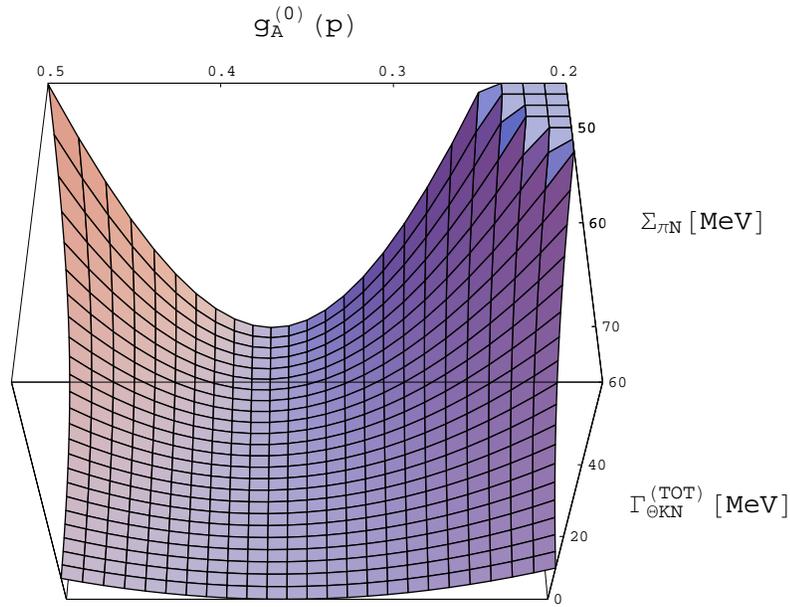}
\caption{The total decay width of $\Theta^+\to K N$ as a function of 
$g_A^{(0)}$ and $\Sigma_{\pi N}$.}
\label{fig:3}
\end{figure}
Figure~\ref{fig:3} shows the results of the total decay width of
the $\Theta^+\to K N$ as a function of $g_A^{(0)}$ and $\Sigma_{\pi
  N}$ in a three dimensional plot.   As shown in Fig.~\ref{fig:3}, the
minimum of the total decay width $\Gamma_{\Theta KN}$ is found to be
around $g_A^{(0)}=0.37$ with $\Sigma_{\pi N}=65$ MeV.  Thus, using
this interrelation among $g_A^{(0)}$, $\Sigma_{\pi N}$, and
$\Gamma_{\Theta KN}$, we can consider a certain window for their
values within the present analysis.  

\begin{figure}[ht]
\centering
\includegraphics[scale=0.6]{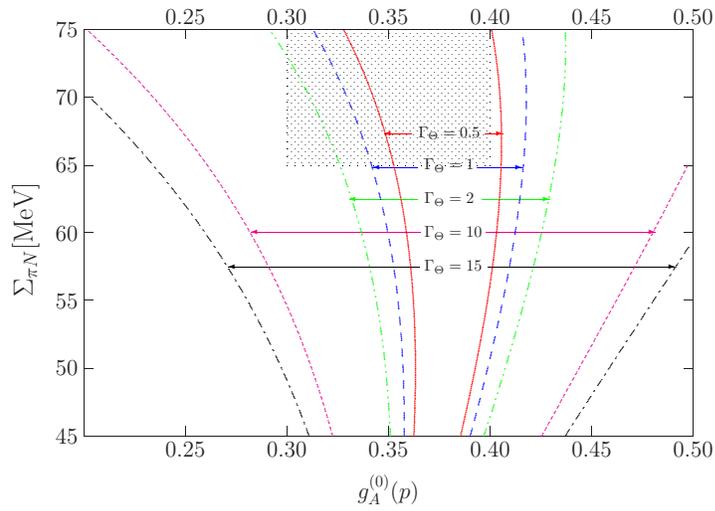}

\caption{The total decay width of $\Theta^+\to K N$ in units of {\rm MeV}
  as a function of $g_{A}^{(0)}$ and $\Sigma_{\pi N}$.  The shaded
  area denotes the ranges of $g_{A}^{(0)}$: $0.3- 0.4$ and of
  $\Sigma_{\pi N}$: $65 - 75$ MeV. }
\label{fig:4}
\end{figure}

Figure~\ref{fig:4} shows the window for the total decay width of
the $\Theta^+\to K N$, given $g_A^{(0)}$ and $\Sigma_{\pi N}$.  
The shaded rectangle indicates the area where one has generally
accepted experimental values of $g_{A}^{(0)}$ and  $\Sigma_{\pi N}$, 
i.e. $0.3- 0.4$ and $65 - 75$ MeV, respectively, 
and simultaneously a $\Gamma_{\Theta KN} \leq 1$ MeV.  It is of great 
interest to see that the range of $g_{A}^{(0)}$ is compatible with a
theoretical investigation~\cite{Wakamatsu:2006ba}, based on the
$\chi$QSM,and on the COMPASS and HERMES measurements of the deuteron
spin-dependent structure function~\cite{Ageev:2005gh,
Alexakhin:2006vx, Airapetian:2006vy}. It is worthwhile to mention that
the values of $g_A^{(0)}$ in the present analysis is almost the
same as theoretical results within the
$\chi$QSM~\cite{Wakamatsu:1999nj,Silva:2005fa}.  The range of
$\Sigma_{\pi N}$ given above is consistent with a recent
analysis~\cite{Schweitzer:2003fg}.  If one interprets the result of
the DIANA collaboration~\cite{Barmin:2006we} as identification of the
$\Theta^+$, namely the formation of a narrow $pK^{0\text{ }}$ peak
with mass of $1537\pm2$ MeV/$c^{2}$ and width of $\Gamma=0.36\pm0.11$
MeV in the $K^{+}n\rightarrow K^{0}p$ transition, then that result is
inside the shaded area of Fig.~\ref{fig:4}.  

\section{Summary and Conclusions}
In the present talk, we have reported the recent investigation on the
magnetic transitions and axial-vector transitions from the baryon
antidecuplet to the nucleons, emphasizing, in particular, the
$\Theta^+$ baryon and the new $N^* (1675)$ resonance from the
GRAAL and Tohoku LNS experiments.  We used the the {\em
  model-independent approach} within the framework 
of the chiral quark-soliton model, thereby taking explicit
SU(3)-symmetry breaking into account.  The parameters in the model
are all fixed by the known experimental data, i.e. octet magnetic
moments, hyperon semileptonic decay constants, singlet axial-vector
constant, octet masses, and the mass of the $\Theta^+$, where the
residual freedom is parametrized by the pion-nucleon sigma term,
$\Sigma_{\pi N}$.  The results for $\mu_{N\Delta}$ and
$\mu_{\Lambda^0\Sigma^0}$ are well reproduced, compared to the
experimental and empirical data.  The transition magnetic moment
$\mu_{\Sigma^-\Sigma^{\ast -}}$, which has only a non-zero value due
to explicit SU(3)-symmetry breaking, is found to be below its upper
bound extracted from the SELEX data~\cite{Molchanov:2004iq}.    

The transition magnetic moment $\mu_{nn_{\overline{10}}^*}$ turns
out to be rather sensitive to the value of $\Sigma_{\pi N}$ due to
the constructive interference of the parameters $w_1(\Sigma_{\pi
  N})$ and $w_1(\Sigma_{\pi N})$.  The value of the
$\mu_{pp_{\overline{10}}^*}$ is rather small in comparison with  
that of the $\mu_{nn_{\overline{10}}^*}$ due to the explicit
SU(3)-symmetry breaking.  As a result, the present predictions for the
transition magnetic moments $\mu_{pp_{\overline{10}}^*}$ and
$\mu_{nn_{\overline{10}}^*}$ are consistent with the recent GRAAL data 
on $\gamma p\rightarrow \eta p$ and $\gamma n\rightarrow \eta n$. This
supports the view that the new resonance $N^*(1675)$ corresponds to a 
neutron-like member of the pentaquark baryon antidecuplet. 

We also presented the recent results of the transition axial-vector
coupling constant for the $\Theta^+\to K^{+} n$, $g_A^{*(\Theta\to
  n)}$ and the total decay width of the $\Theta\to KN$.  We showed
that the $g_A^{*(\Theta\to n)}$ decreases as $g_{A}^{(0)}$
increases. It also depends on the $\pi N$ sigma term in such a way
that it is getting smaller as the $\Sigma_{\pi N}$ increases.  It was
also shown that the $g_A^{*(\Theta\to n)}$ turns out to be 
negative around $g_{A}^{(0)}\simeq 0.37$.

The total width $\Gamma_{\Theta KN}$ of the $\Theta^+\to KN$ decay
was finally investigated.  Since it is proportional to the square
of the transition axial-vector constant $g_A^{*(\Theta\to n)}$, it
is rather sensitive to the $g_A^{*(\Theta\to n)}$.  The
$\Gamma_{\Theta KN}$ is getting suppressed as the singlet
axial-vector constant $g_{A}^{(0)}$ increases.  However, since the
$g_A^{*(\Theta\to n)}$ becomes negative around 0.37, the 
$\Gamma_{\Theta KN}$ starts to increase around 0.37. As a result, 
the total decay width $\Gamma_{\Theta KN}$ turns out to be smaller
than 1 MeV for values of the $g_{A}^{(0)}$ and $\Sigma_{\pi N}$
larger than 0.31 and 65 MeV, respectively.

According to the analysis of the total width $\Gamma_{\Theta KN}$, we 
draw a conclusion as follows: The known data of semileptonic decays 
combined with $0.3 \leq  g_{A}^{(0)} \leq 0.4$ and $\Sigma_{\pi N}
\geq 65$ MeV is compatible with the existence of a $\Theta^+$
pentaquark having a small width of the total decay $\Theta^+\to KN$,
that is $\Gamma_{\Theta KN}\le 1~ \rm{MeV}$.  

\section*{Acknowledgments}
One of the authors (H.-Ch.K.) thanks the Yukawa Institute for
Theoretical Physics at Kyoto University, where this work was completed
during the YKIS2006 on "New Frontiers on QCD".  The present work is
supported by the Korea Research Foundation Grant 
funded by the Korean Government(MOEHRD) (KRF-2006-312-C00507).  The
work of K.G. and Gh-S.Y. is partially supported by the
Transregio-Sonderforschungsbereich Bonn-Bochum-Giessen as well as by 
the Verbundforschung and the International Office of the Federal
Ministry for Education and Research (BMBF).

\end{document}